\RequirePackage{etoolbox}
\patchcmd{\bibliographystyle}{#1}{naturemagnourlnoarxiv}{}{}

\documentclass[10pt]{wlscirep}
\usepackage[utf8]{inputenc}
\usepackage[T1]{fontenc}

\usepackage{bm}
\usepackage{graphicx} 
\usepackage{float}    
\usepackage{subfig}   

\usepackage{amsmath}
\usepackage{amssymb}
\usepackage[mathscr]{euscript} 

\newsavebox{\foobox}
\newcommand{\slantbox}[2][0]{\mbox{
        \sbox{\foobox}{#2}%
        \hskip\wd\foobox
        \pdfsave
        \pdfsetmatrix{1 0 #1 1}%
        \llap{\usebox{\foobox}}%
        \pdfrestore
}}
\newcommand\unslant[2][-.15]{\slantbox[#1]{$#2$}}
\newcommand{\micron}{\mbox{$\unslant\mu$\hspace{-1pt}m}}

\usepackage{fancyhdr}
\newcommand{\SI}[1]{Sec.~\ref{#1}:~\nameref{#1}}

\title{Ultra-high-$\bm{Q}$ resonances in plasmonic metasurfaces}

\author[1,$\dagger$]{M.~Saad Bin-Alam}
\author[2,*,$\dagger$]{Orad~Reshef}
\author[3,1]{Yaryna~Mamchur}
\author[2]{M.~Zahirul~Alam}
\author[4]{Graham~Carlow}
\author[2]{Jeremy~Upham}
\author[4]{Brian~T.~Sullivan}
\author[2]{Jean-Michel~M\'{e}nard}
\author[5]{Mikko~J.~Huttunen}
\author[1,2,6]{Robert~W.~Boyd}
\author[1,2]{Ksenia~Dolgaleva}

\affil[1]{School of Electrical Engineering and Computer Science, University of Ottawa, Ottawa, ON K1N~6N5, Canada}
\affil[2]{Department of Physics, University of Ottawa, 25 Templeton Street, Ottawa, ON K1N 6N5, Canada}
\affil[3]{National Technical University of Ukraine ``Igor Sikorsky Kyiv Polytechnic Institute,'' Kyiv, Ukraine}
\affil[4]{Iridian Spectral Technologies Inc., 2700 Swansea Crescent, Ottawa, ON K1G 6R8, Canada}
\affil[5]{Photonics Laboratory, Physics Unit, Tampere University, P.O. Box 692, FI-33014 Tampere, Finland}
\affil[6]{Institute of Optics and Department of Physics and Astronomy, University of Rochester, Rochester, NY 14627, USA}
\affil[*]{Corresponding author: orad@reshef.ca}
\affil[$\dagger$]{MSBA and OR contributed equally to this work.}

\begin{abstract}Plasmonic nanostructures hold promise for the realization of ultra-thin sub-wavelength devices, reducing power operating thresholds and enabling nonlinear optical functionality in metasurfaces. However, this promise is substantially undercut by absorption introduced by resistive losses, causing the metasurface community to turn away from plasmonics in favour of alternative material platforms (e.g., dielectrics) that provide weaker field enhancement, but more tolerable losses. Here, we report a plasmonic metasurface with a quality-factor (Q-factor) of 2340 in the telecommunication C band by exploiting surface lattice resonances (SLRs), exceeding the record by an order of magnitude. Additionally, we show that SLRs retain many of the same benefits as localized plasmonic resonances, such as field enhancement and strong confinement of light along the metal surface. Our results demonstrate that SLRs provide an exciting and unexplored method to tailor incident light fields, and could pave the way to flexible wavelength-scale devices for any optical resonating application.
\end{abstract}

\begin{document}

\flushbottom
\maketitle

\section{Introduction}
Metallic nanostructures are essential to many applications in photonics, including biosensing~\cite{Anker2008}, spectroscopy~\cite{Nie1997,Willets2007}, nanolasing~\cite{Azzam2020}, all-optical switching~\cite{Zheludev2011},nonlinear optical processes~\cite{Kauranen2012}, and metasurface technologies~\cite{Yu2011,Won2017,Meinzer2014}. 
These plasmonic elements form flexible components with geometry-dependent responses and have many desirable properties, such as the possibility to confine light to sub-wavelength scales, and large local-field enhancements~\cite{Meinzer2014,Maier2007}. Metals also possess intrinsic nonlinear optical constants that are many orders of magnitude larger than dielectric materials~\cite{Boyd2020}. 

When structured at the sub-wavelength scale~\cite{Won2017,Meinzer2014,Oldenberg1998}, individual nanostructures exhibit localized surface plasmon resonances (LSPRs), where electromagnetic fields couple to the free-electron plasma of a conductor at a metal-dielectric interface~\cite{Maier2007, Kauranen2012}. Depending on its shape, an individual nanoparticle may be polarized by an incident light beam, acting as a lossy dipole antenna~\cite{Novotny2011} and trapping light for a short period of time. 
In contrast to other photonic resonant devices such as whispering gallery mode resonators, microring resonators or photonic crystals~\cite{Zhang2010b, Ji2017, Asano2017}, 
resonating dipoles in a metasurface can easily be accessed by a beam propagating in free space, and require only a sub-wavelength propagation region for operation. Therefore, a plasmonic metasurface resonator enables a series of specialized optical responses, including phase-matching-free nonlinear optical effects~\cite{Kauranen2012,krasnok2018nonlinear}, strongly localized field enhancements~\cite{Meinzer2014}, multi-mode operation~\cite{Celebrano2015}, and a spatially localized optical response~\cite{Yu2011}. Such a metasurface with a large ${Q}$-factor could be used as a cavity for applications that need increased light-matter interactions, small mode volumes, large field enhancements and large optical nonlinearities, such as an ultra-flat nano-laser with a large transverse mode size~\cite{Zhou2013,Azzam2020} or frequency conversion applications ({\emph{e.g.},} nonlinear harmonic generation~\cite{Michaeli2017a}, or THz-wave generation~\cite{Luo2014b}). One frequently cited limitation of LSPR-based metasurfaces are their low $Q$-factors (\emph{e.g.}, $Q<10$) due to the intrinsic Ohmic losses present in metals at optical frequencies~\cite{Maier2007,Choi2018,Sain2019,Koshelev2020}. As the $Q$-factor is related to the light-matter interaction time as well as to enhancements to the electric field, it is typically desirable to maximize this quantity~\cite{Zhang2010b}. Low $Q$-factors therefore make many potential applications of plasmonics-based metasurface devices impractical, and new methods for obtaining large $Q$-factor resonances in a metasurface have long been sought after.

\begin{figure}[htb]
\centering
\includegraphics[width=115mm]{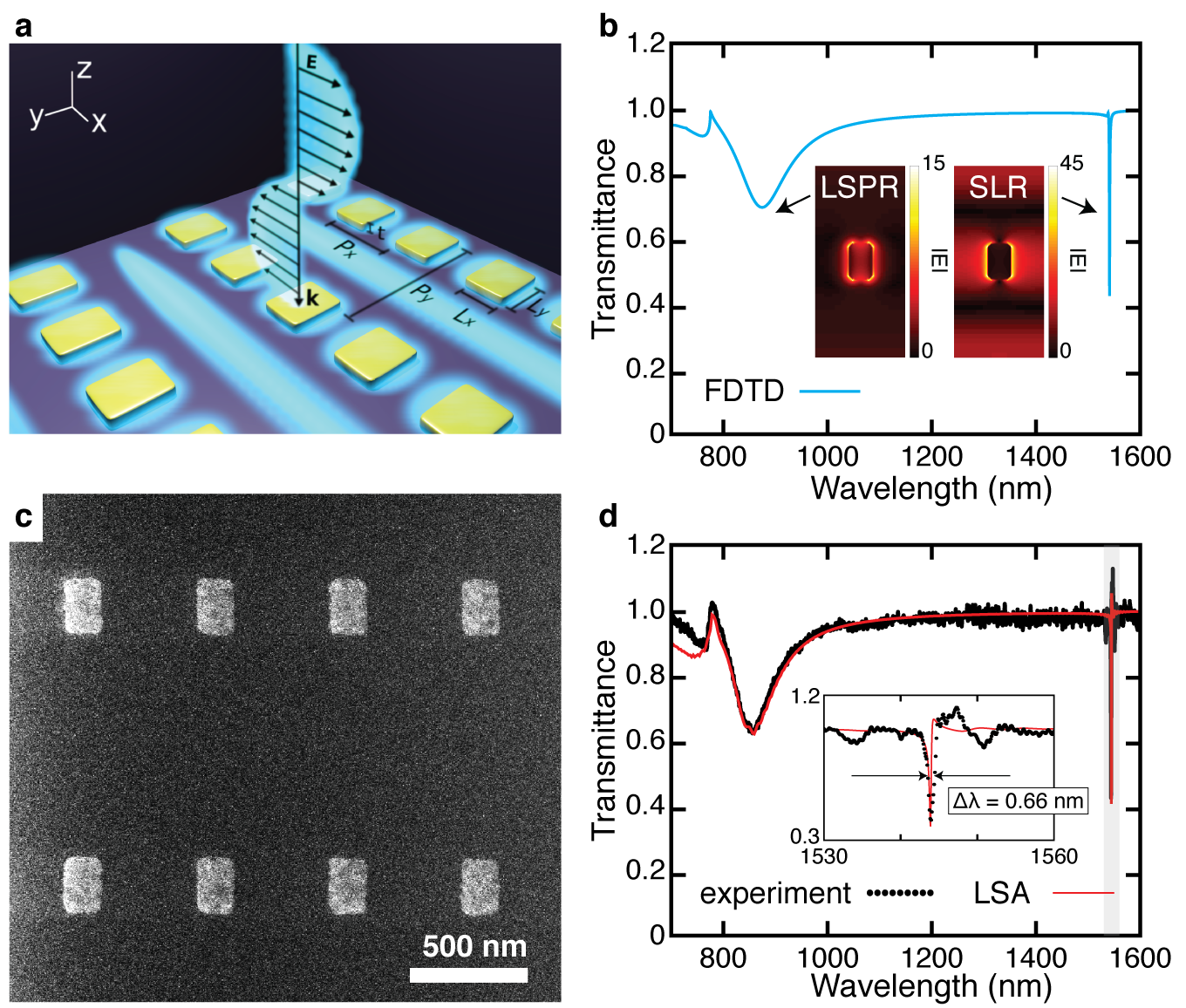}
\caption{{\bf~|~High-$\bm{Q}$ metasurface nanocavities using arrays of plasmonic nanostructures. a,~} Schematic of the metasurface consisting of a rectangular array of rectangular gold nanostructures. Here, $L_x=130$~nm, $L_y=200$~nm, $t=20$~nm, $P_x=500$~nm, and $P_y=1060$~nm. The blue shaded regions illustrate the electric field, reproducing the mode structure in the inset of {\bf (b)}. {\bf b,~}Numerical (FDTD) calculations of the transmission spectrum of this metasurface for $x$-polarized light. Both the LSPR and the SLR are observed in these results. {\bf Inset,} The simulated magnitude of the electric field $|E|$ for the entire unit cell of both LSPR and SLR modes in the $x$-$y$ plane that bisects the nanoparticles. The colorbar indicates the relative magnitude when normalized to the incident plane wave. {\bf c,~}Helium ion microscope image of the fabricated metasurface prior to cladding deposition. {\bf d,~}Measured transmission spectrum (black dots) and fits to semi-analytic calculations (LSA, red line). {\bf Inset,~}Zoomed plot of the highlighted region in {\bf (d)}. Fitting the measurement to a Lorentzian function yields a linewidth of $\Delta \lambda=0.66$~nm, corresponding to $Q=2340$ (see \SI{SEC:fit}).}
\label{fig:schematic}
\end{figure}

The optical response of coupled plasmonic nanoresonators has been a topic of intense study~\cite{Nordlander2004}. Notably, plasmonic metasurfaces of large periodically arranged nanostructures support collective resonances called surface lattice resonances (SLRs)~\cite{Auguie2008,chu2008experimental,kravets2008extremely,vecchi2009surface,Kravets2018,Khlopin2017,Zou2004}. Here, the individual responses from the surface plasmons of many individual nanostructures form a collective response that couples to in-plane diffraction orders of the periodic array~\cite{Auguie2008, Kravets2018}. As a consequence, a relatively high-$Q$ resonance can emerge at an optical wavelength $\lambda_ \mathrm{SLR}\approx nP$, close to the product of the refractive index of the background medium $n$ and the lattice period $P$~~\cite{Zou2004,Auguie2008}. Recent theoretical studies of this platform have predicted $Q$-factors on the order of $10^3$ by properly engineering the dimensions of the individual nanostructures and the period of the lattice~\cite{Zou2004,Zakomirnyi2017,Khlopin2017}, hinting at the possibility of combining the aforementioned benefits of metals with long interaction times provided by high $Q$-factors.  However, to date, the highest experimentally observed $Q$-factor in an SLR-based metasurface is 430~\cite{Deng2020}. The disparity between theory and experiment has been attributed to \a variety of reasons, including poor spatial coherence of light beams~\cite{kravets2008extremely, Li2014b}, small array sizes~\cite{Kravets2018, Rodriguez2012, Khlopin2017} fabrication imperfections~\cite{Khlopin2017, Kravets2018}, and the addition of an adhesion layer~\cite{Le-Van2019}. 

Inspired by this discrepancy, here we perform a detailed investigation to determine the dominant factors that most drastically affect the observed $Q$ of an SLR-based metasurface: the nanostructure geometry, the array size, and the spatial coherence of the probing light source. Using the results of this study, we demonstrate a plasmonic metasurface capable of supporting ultra-high-$Q$ SLRs.

\section{Results}
The metasurface in consideration consists of a rectangular array of rectangular gold nanostructures embedded in a homogeneous silica glass ($n\sim1.45$) environment (Fig.~\ref{fig:schematic}a). The lattice constant $P_y=1060$~nm was selected to place the SLR wavelength in the telecommunication window; $P_x=500$~nm was reduced from a square lattice, increasing the nanoparticle density and consequently increasing the extinction ratio of the resonance. The overcladding is carefully matched to the substrate material to ensure a symmetric cladding index, as it has been shown that the $Q$ of an SLR may be affected by the homogeneity of the environment~\cite{auguie2010diffractive,Auguie2008,Orad2019}. As shown by the numerical predictions in Fig.~\ref{fig:schematic}b, for an $x$-polarized beam, this metasurface is expected to support an LSPR at $\lambda_\mathrm{LSPR}=830$~nm and an SLR of the first type around $\lambda_\mathrm{SLR}=1550$~nm (See \SI{SEC:SLR_Type}). The SLR linewidth is substantially narrower than that of the LSPR, corresponding to a much higher $Q$-factor. Incidentally, the inset field profiles in Fig.~\ref{fig:schematic}b also reveal that the SLR provides a more significant field enhancement, with $|E_\mathrm{max}(\lambda_\mathrm{SLR})| \sim 3|E_\mathrm{max}(\lambda_\mathrm{LSPR})|$. Figure~\ref{fig:schematic}c shows an image of the fabricated device with dimensions matching those of the simulations. The measured transmission spectra are presented in Figs.~\ref{fig:schematic}d, closely matching the predicted spectrum. Notably, the full width at half-maximum of the linewidth is only $\Delta \lambda=0.66$~nm, corresponding to a quality factor of $Q=2340$. This value exceeds the record for plasmonic metasurfaces by an order of magnitude~\cite{Thackray2015,Le-Van2019, Deng2020}, and is among the highest reported in a metasurface. It is roughly within a factor of two of semi-analytic calculations performed using the Lattice Sum Approach (LSA), where $Q\sim 5000$ (see Methods for details). In order to observe this value for the $Q$-factor, both the metasurface and the measurement apparatus needed to be arranged with a few considerations in mind which we describe in greater detail below.

\subsection{The role of nanoparticle polarizability}
First, the individual structures need to be engineered to exhibit the appropriate response at $\lambda_\mathrm{SLR}$. The optical response of a nanostructure can be approximated using the polarizability of a Lorentzian dipole,
\begin{align}
    \alpha(\omega)=\frac{A_0}{\omega-\omega_0+i\gamma},\label{EQ:alpha-omega}
\end{align}
where $A_0$ is the oscillator strength, $\omega_0=2\pi c / \lambda_\mathrm{LSPR}$ corresponds to the nanoparticle resonance frequency, and $\gamma$ is the damping term. These quantities all depend on the particle geometry~\cite{Oldenberg1998} (here, the length $L_y$ and width $L_x$ of a rectangular bar).
The contribution of the particle lattice to the polarizability can be introduced using the lattice-sum approach~\cite{Markel2005,Orad2019}:
\begin{align}
    \alpha^*(\omega) = \frac{\alpha(\omega)}{1-\epsilon_0\alpha(\omega)S(\omega)}, \label{EQ:alpha_star}
\end{align}
where $\alpha^*(\omega)$ is known as the effective polarizability of the entire metasurface, and $S(\omega)$ corresponds to the lattice sum. This latter term depends only on the arrangement of the lattice. An SLR appears approximately where $S(\omega)$ exhibits a pole, at $\omega_\mathrm{SLR}=(2\pi c / \lambda_\mathrm{SLR})$. At this spectral location, the individual responses of all of the nanostructures contribute cooperatively~\cite{Markel2005}.

\begin{figure}[!b]
\centering
\includegraphics[width=92.5mm]{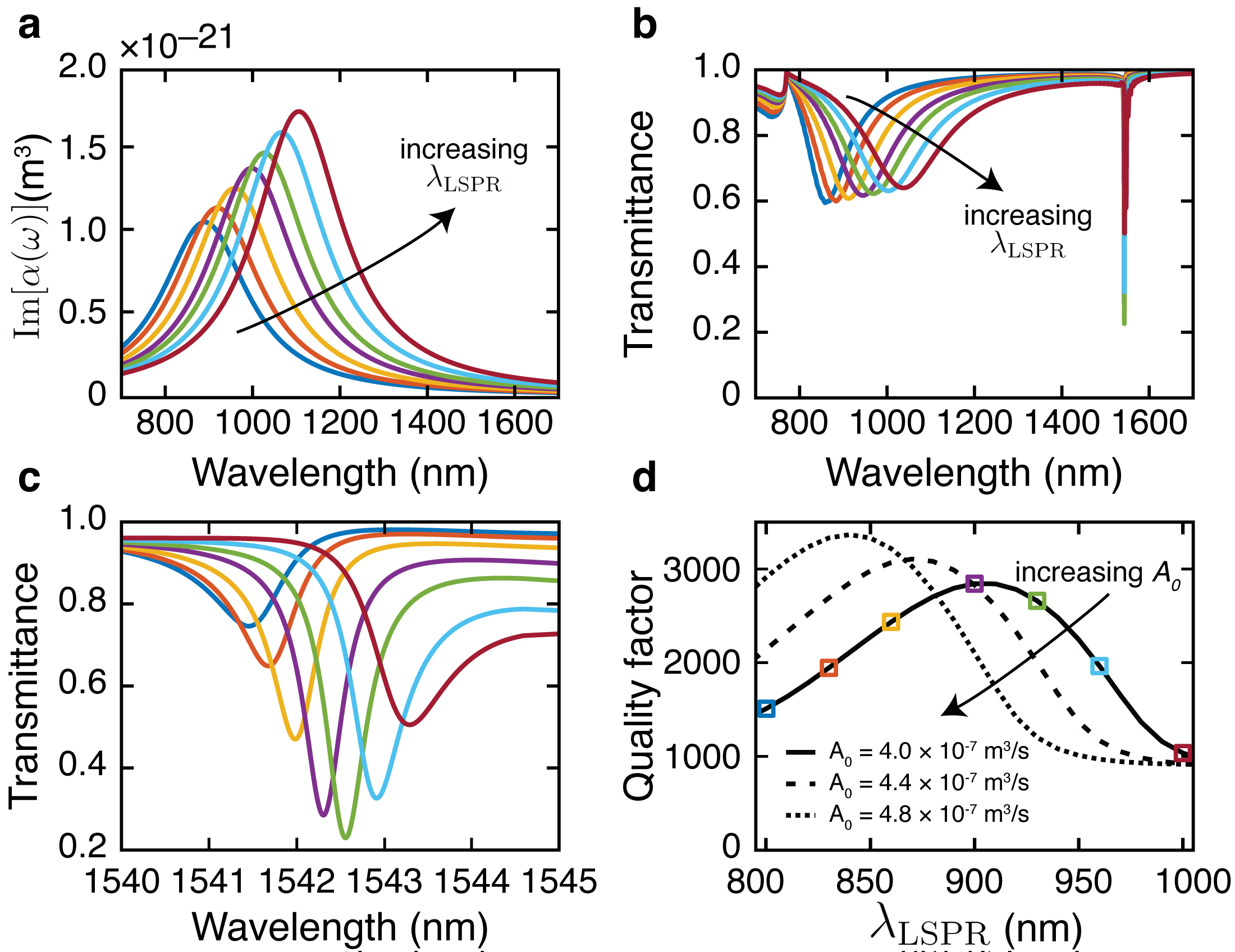}
\caption{{\bf~|~Coupling to a surface lattice resonance.} The colors in parts {\bf (a)} through {\bf (d)} are consistent, corresponding to the same type of nanoparticle. {\bf a,~}The imaginary part of the individual particle polarizability for various nanostructures with increasing resonance wavelength $\lambda_\mathrm{LSPR}$, holding both the oscillator strength $A_0$ and the damping term $\gamma$ fixed. Here, $\lambda_\mathrm{LSPR}$ is tuned from 800~nm to 1000~nm. {\bf b,~}Simulated broadband transmission spectra for gold nanostructure arrays as a function of tuning $\lambda_\mathrm{LSPR}$. By tuning the LSPR wavelength, the extinction factor of the SLR is observed to change near $\lambda=1542$~nm. While $\lambda_\mathrm{LSPR}$ changes dramatically, the SLR wavelength $\lambda_\mathrm{SLR}$ does not change much. {\bf c,~}Zoomed in plot of the SLR in \textbf{(b)}. {\bf d,~}The $Q$-factor of the surface lattice resonance as a function of $\lambda_\mathrm{LSPR}$ for various oscillator strengths $A_0$. The optimal LSPR wavelength for a high-$Q$ SLR changes as a function of $A_0$. The oscillator strength $A_0$ increases from $3.98\times10^{-7}$ to $4.77\times10^{-7}\mathrm{~m}^3/\mathrm{s}$, roughly corresponding to a 20\% increase in the particle volume. The squares indicate the $Q$ values extracted from the curves in \textbf{(c)}.}
\label{FIG:CriticalCoupling}
\end{figure}

Equation~\eqref{EQ:alpha_star} may be used to predict the optical response of the entire metasurface, including the behaviour of its many resonances, as a function of the geometry of its nanostructures (see Methods): by changing the geometry of a nanostructure~\cite{Oldenberg1998,teperik2012design}, its individual resonance wavelength $\lambda_\mathrm{LSPR}$, oscillator strength $A_0$, and damping constant $\gamma$ are all modified. In turn, adjusting these values changes the polarizability of the nanostructures throughout the spectrum, including at the surface lattice resonance wavelength $\alpha(\omega_\mathrm{SLR})$, and therefore also the response of the entire metasurface at this wavelength $\alpha^*(\omega_\mathrm{SLR})$. Here, we adjust the above parameters by changing the dimensions of the nanostructures (see Methods), while the parameters could be alternatively modified by considering altogether different nanostructure shapes, such as nanorings, nanorods or core-shell nanoparticles
~\cite{teperik2012design}. By contrast, the spectral location of the SLR wavelength is dictated mainly by the lattice period and the background index $\lambda_\mathrm{SLR}\approx nP$~\cite{Zou2004,offermans2011universal,Huttunen2016}. In other words, the lattice configuration governs the presence of the SLR, and the nanostructure geometry dictates its coupling efficiency to free space. Indeed, recent theoretical studies in this platform have shown $Q$-factors on the order of $10^3$ by properly selecting the dimensions of the individual nanostructures~\cite{Zakomirnyi2017,Khlopin2017}.

We reproduce this dependence in this platform explicitly by plotting the calculated transmission of a metasurface (see Methods) as a function of nanostructure resonance wavelength $\lambda_\mathrm{LSPR}$ (Fig.~\ref{FIG:CriticalCoupling}). (The dependence of the SLR behaviour on particle dimensions, which is connected to the resonance wavelength, is also demonstrated using full-wave simulations in \SI{SEC:particle_dimension_Sweep}.) Here, we hold the oscillator strength $A_0$ and damping term $\gamma$ constant and slowly increase the nanoparticle resonance wavelength $\lambda_\mathrm{LSPR}$. Note that the resonance position differs slightly from the position of the dip due to the incorporation of a long-wavelength correction~\cite{Jensen1999}. In Figs.~\ref{FIG:CriticalCoupling}b~--~c, the SLR wavelength does not change substantially from its location around $\lambda_\mathrm{SLR}=1542$~nm; however, the extinction ratio $\Delta T$ and the linewidth $\Delta \lambda$ of the resonance change dramatically. In Fig.~\ref{FIG:CriticalCoupling}d, we plot the extracted $Q$-factors for these SLRs, and for other values of $A_0$, as well (see \SI{SEC:fit} for the fits). Based on well-established relationships between nanoparticle geometry and polarizability~\cite{Huang2009, Maier2007} this $A_0$ range corresponds to a change in nanoparticle volume of roughly 20\%. We find that for every given value of $A_0$, there is a corresponding $\lambda_\mathrm{LSPR}$ for which light couples optimally to the lattice resonance at $\lambda_\mathrm{SLR}$ and produces the highest $Q$-factor. The optimal conditions are therefore found in the balance between increasing $\alpha$ relative to $P_y$ (\emph{i.e.,} increasing coupling strength), and maintaining a large spectral gap between $\lambda_\mathrm{LSPR}$ and $\lambda_\mathrm{SLR}$ (\emph{i.e.,} limiting Ohmic losses associated with metallic nanoparticles). The trade-off between coupling and loss is a traditional one for optical resonators and is reproduced in the SLR-based metasurface platform~\cite{McKinnon2009}.

\subsection{Effect of array size}
Next, we study the dependence of the $Q$-factor on the array size. For certain metasurfaces, it has already been predicted that larger array sizes lead to better device performance~\cite{Rodriguez2012,Zundel2019}. This dependence makes some intuitive sense ---  since high-$Q$ operation requires low absorption losses, we are required to operate the device far from the LSPR. However, at a sufficiently far operating wavelength, the scattering cross-section is also small, resulting in each antenna scattering very weakly. Consequently, far from the LSPR, one requires a sufficiently large number of scatterers to build up the resonance. Equivalently, the standing wave mode in an SLR consists of counter-propagating surface waves; therefore, a larger array provides an expanded propagation length in the cavity to support these modes.

\begin{figure*}[htb]
\centering
{\includegraphics[width=89mm]{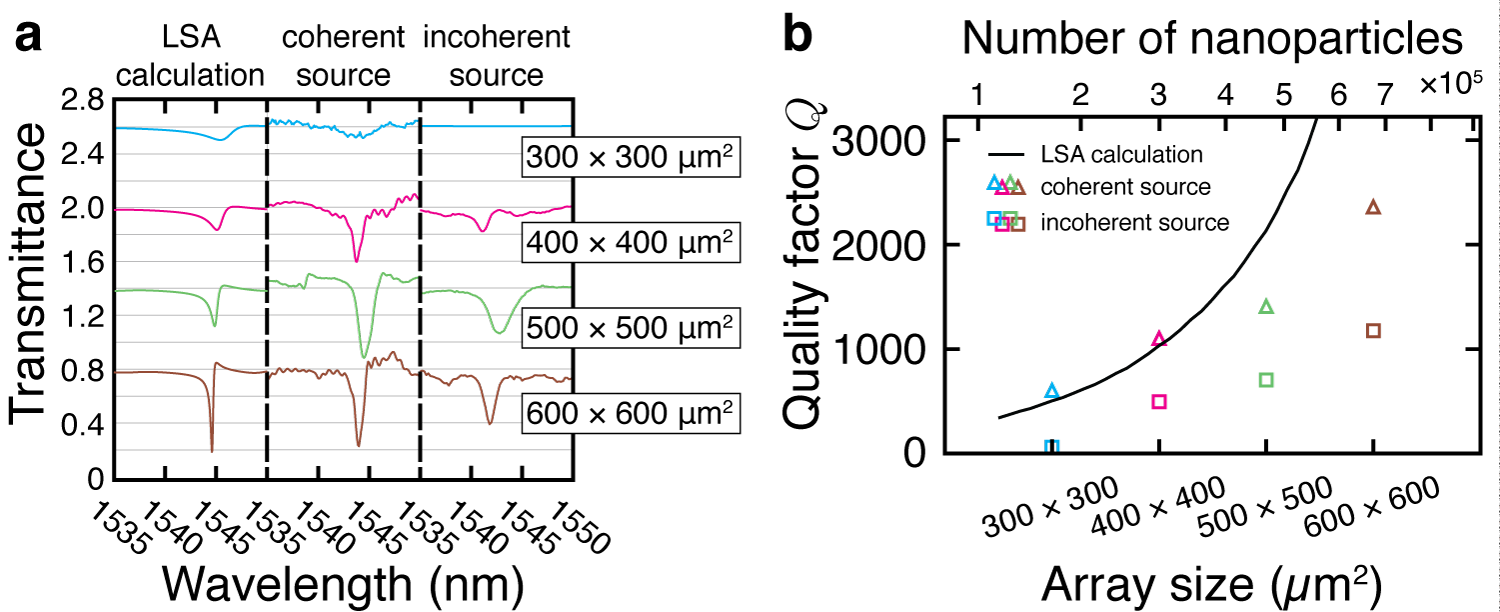}}
\caption{{\bf~|~Effects of array size and spatial coherence of light source.} {\bf a,~}Calculated and measured (using coherent and incoherent sources) transmission spectra for identical metasurface arrays of varying size (from top to bottom: $300\times300$, $400\times400$, $500\times500$, and $600\times600$~\micron$^2$, respectively). The spectra are offset for clarity, and each vertical division corresponds to an increment of $\Delta T=0.2$ in transmittance. 
{\bf b,~} The $Q$-factors extracted from Lorentzian fits to the calculations and to the measurements shown in {\bf (a)}. An increase in the number of nanostructures in the array results in an increase in the estimated $Q$-factors. Additionally, the observed $Q$-factor is globally larger for each array when measured using the coherent source.}
\label{FIG:ArraySize}
\end{figure*}

To examine the dependence of $Q$ on the number of nanostructures explicitly, we fabricated and characterized a series of devices of increasing array size. Figure~\ref{FIG:ArraySize} shows the resulting transmission spectra, as well as their corresponding semi-analytic predictions. The observed $Q$-factors increase monotonically as a function of array size (Fig.~\ref{FIG:ArraySize}b -- see \SI{SEC:fit} for the fits). In the smallest array ($300\times300$~\micron$^2$), the SLR is almost imperceptible. This trend might help explain the relatively low $Q$ values observed in previous studies~\cite{Meinzer2014,Kravets2018,Khlopin2017,Rodriguez2012} where array sizes were typically no larger than 250$\times$250 $\mu$m$^2$, likely due to the relatively slow write-speed of the electron-beam lithography process necessary for fabrication~\cite{Auguie2008,Le-Van2019}. By contrast, our devices have array sizes reaching up to $600\times600$~\micron$^2$ (see \SI{SEC:device_image}). 

\subsection{The role of spatial coherence}
Finally, it is of critical importance to consider all aspects of the characterization system in order to get an accurate measurement of the $Q$-factor. In particular, we have found that the spatial coherence of the probe beam was critical to obtaining a clean measurement of the dip in transmission indicating a resonance. A spatially coherent beam, such as a laser, excites every region of the metasurface in phase, producing a resonance feature that is both deeper and narrower compared to using spatially incoherent source. Additionally, the higher-order modes of the lattice are more sensitive to angular variance in the measurements, leading to broader peaks when using incoherent sources~\cite{Li2014b}. Furthermore, in our particular experiment, the transmitted signal from our coherent supercontinuum source was both brighter and could also be better collimated than our incoherent thermal source. Therefore, the light collected from the metasurface array could be isolated with a smaller pinhole in the image plane, selecting the signal coming from nanostructures at the centre of the array with a more uniform collective response.

In Fig.~\ref{FIG:ArraySize}, we compare the performance of the metasurface when illuminated using different light sources: a broadband supercontinuum laser (\emph{i.e.,} a well-collimated coherent source), and a tungsten-halogen lamp. The comparison between these measurements indicates that the $Q$ increases with the coherence of the light source --- using the thermal light source reduces the $Q$-factor by a factor of 2--5 when compared to the laser. Additionally, it decreases the resonance coupling strength, as is evident from the reduced extinction ratio of the SLRs. Figure~\ref{FIG:ArraySize}b summarizes the $Q$-factors extracted from these measurements and compares them to numerical predictions. LSA calculations predict that $Q$-factors increase as a function of array size; this trend continues for both smaller and larger devices than those probed experimentally. Note that even when using an incoherent source, the largest array still produces a very large $Q$-factor ($Q\sim 1000$). The observation of such a high $Q$ using an incoherent source reinforces the validity of our aforementioned metasurface design criteria --- that is, the importance of the choice of nanostructure geometry and of the array size.

In some of the measurements, the value for the normalized transmittance can be seen to exceed unity (\emph{i.e.,} $T>1$). We speculate that this is because the nanostructures aid in coupling to the substrate, reducing the reflections from the first interface.

\section{Discussion}

Despite promising results, Fig.~\ref{FIG:ArraySize}b also highlights some discrepancies between the simulation and the experiment for the largest arrays, notably reducing the measured $Q$-factors. This disparity could be due to multiple reasons, which we enumerate below. First, the prediction produced by the LSA might be overestimating the $Q$ by assuming that each nanoparticle is excited with a constant-valued local field. This assumption cannot be entirely correct for a Gaussian beam and a finite array, where particles closer to the boundaries of the array feel a weaker local field than the particles near the center. Secondly, the fabrication procedure produces stitching errors which become more important for larger arrays. This added disorder might contribute to the reduction in $Q$. Lastly, the $Q$-factors might be limited due to additional measurement considerations, such as the finite coherence length of the light source, or imperfections with the collimation.

In this work we only looked into rectangular nanoparticles in rectangular lattices. Based on LSA calculations and the discrete-dipole approximation (DDA) used in previous work~\cite{Auguie2008,Huttunen2016,Huttunen2018,Huttunen2019, Orad2019}, it is evident that any particle geometry (\emph{e.g.,} cylindrical, rectangular or, triangular) that can be approximated by dipoles with the same Lorentzian parameters $A_0$, $\lambda_\mathrm{LSPR}$, and $\gamma$ will yield an identical SLR $Q$-factor. For nanoparticles that cannot be modeled by dipoles --- regardless of the particle geometry --- the SLR $Q$-factor will be the same provided that the polarizability at $\lambda_\mathrm{SLR}$ remains the same. Regarding different lattice configurations, the spectral responses of other lattice geometries such as hexagonal, orthorhombic, Kagome, are likely to be different than the rectangular lattice design we have adopted. However, lattice sums can be computed for these regular lattices, and therefore they can also be treated using our method. Therefore, strategies presented in this work are largely blind to the specific lattice arrangements, and its conclusions will be helpful in obtaining resonances with large-Q factors in other geometries.

The $Q$-factors for the type of device presented here could be further increased, however, by considering larger arrays, or by further optimizing the nanostructure dimensions --- instead of rectangles, a more intricate nanostructure shape could tailor the Lorentzian dipole coefficients $A_0$, $\lambda_\mathrm{LSPR}$, and $\gamma$ more independently to allow for optimal coupling and higher extinction ratios. These shapes include L-shaped antennas~\cite{husu2010particle}, split-ring resonators~\cite{Corrigan2008}, and others that also exhibit higher order moments~\cite{barnes2016particle, alaee2018electromagnetic}. Alternatively, a nanoparticle with a large aspect ratio could increase coupling to more neighbouring particles using out-of-plane oscillations~\cite{Huttunen2016}. Finally, the metasurface shown here can be combined with other established methods to enable multiple simultaneous resonances~\cite{Orad2019,Baur2018,Huttunen2019}.

Table~\ref{TAB:Literature1} contains a short survey of the literature on metasurface nanocavities. Other than the reported $Q$-factors, we have included, when available, information that is relevant to compare their work against ours, such as the operating wavelength, the material platform, the array size and the type of light source used. Our work demonstrates the highest $Q$ by an order of magnitude among metasurfaces with plasmonic components, and is exceeded only by metasurfaces that incorporate a bound-state in the continuum (BIC).
\begin{table}[H]
\centering
\begin{tabular}{|l|c|c|c|c|c|c|}
\hline
Mechanism & $Q$ & $\lambda$ (nm) & Material & Light source & Array size (\micron$^2$) & Reference \\ \hline
 {\bf SLR} & {\bf 2340} & {\bf 1550} & {\bf Au NPs} &  {\bf Supercontinuum} & {\bf 600$\times$600} & {\bf This work} \\ \hline
 LSPR & $<$10 & 700 & Au NPs & Tungs.-Halogen lamp & 3000$\times$3000 & \cite{Rodriguez2011} \\ \hline
  SLR & 25 & 930  & Au NPs & Collimated source & 135$\times$135 & \cite{chu2008experimental} \\ \hline
 SLR & 30 & 850  & Au NPs & Tungs.-Halogen lamp & 3000$\times$3000 & \cite{Rodriguez2011} \\ \hline
 SLR & 40--60 & 600  & Au NPs & Ellipsometer & 200$\times$200 & \cite{kravets2008extremely} \\ \hline
 SLR & 60 & 800  & Au NPs & Tungs.-Halogen lamp & 35$\times$35 & \cite{Auguie2008} \\ \hline
 SLR & 150 & 764  & Au NPs & Tungs.-Halogen lamp & N/A & \cite{Kravets2010} \\ \hline
 SLR & 230 & 900 & Au NPs & Tungs.-Halogen lamp & $\sim 10000\times10000$ & \cite{Yang2015} \\ \hline
 SLR & 300 & 1500  & Au nanostripes & Tungs.-Halogen lamp & 300$\times$100 & \cite{Thackray2015} \\ \hline
 SLR & 330 & 648  & Ag NPs & Tungs.-Halogen lamp & 2500$\times$2500 & \cite{Le-Van2019} \\ \hline
 SLR & 430 & 860  & Au NPs & Tungs.-Halogen lamp & 1000$\times$1000 & \cite{Deng2020} \\ \hline
 Mirror Image & 200 & 5000 & ITO nanorods & Collimated source & N/A  & \cite{Li2014b} \\ \hline
 EIT & 483 & 1380 & Si & Tungs.-Halogen lamp & 225$\times$240 & \cite{Yang2014} \\ \hline
 Fano Resonance & 65 & THz & Al Particles & THz laser & $10000\times10000$ & \cite{Singh2014} \\ \hline
 Fano Resonance & 100 & THz & Au Assym. NPs & FTIR & 150$\times$150 & \cite{Altug2011} \\ \hline
  Fano Resonance & 350 & 1000 & Si & N/A & N/A & \cite{Brener2016} \\ \hline
 Fano Resonance & 600 & 1000 & GaAs & N/A & N/A & \cite{Brener2016} \\ \hline
 BIC & 2750 & 825 & GaAs & Laser & $60\times108$ & \cite{Ha2018} \\ \hline
 Quasi-BIC & 18511 & 1588.4 & Si & Laser & $15\times15$ & \cite{Liu2019} \\ \hline
 
\end{tabular}
\caption{Summary of experimentally obtained $Q$-factors in metasurfaces. The results presented in this work are in bold. $Q$, quality-factor; $\lambda$, resonance wavelength; NP, nanoparticle; SLR, surface lattice resonance; LSPR, localized surface plasmon resonance; EIT, electromagnetically induced transparency, BIC, bound-state in the continuum}\label{TAB:Literature1}
\end{table}

To summarize, we have fabricated and experimentally demonstrated a plasmonic metasurface nanoresonator with a high $Q$-factor which is in excellent agreement with numerical predictions. Our work presents the experimental demonstration of a high-$Q$ plasmonic metasurface nanoresonator with an order-of-magnitude improvement over prior art (see Table~\ref{TAB:Literature1}).
We have found that the observed $Q$-factor obtained from an SLR may be limited by a poor choice of nanostructure dimensions, a small array size, or poor spatial coherence of the source illumination; we hypothesize that one or many of these factors may have been the cause for the low $Q$-factors reported in previous experiments featuring SLRs. Additionally, our device follows simple design principles that can be easily expanded upon to enable multiple resonances to fully tailor the transmission spectrum of a wavelength-scale surface. Our result highlights the potential of SLR-based metasurfaces, and expands the capabilities of plasmonic nanoparticles for many optical applications.

\section*{Methods}
\subsection*{Simulations}
\paragraph*{FDTD:}Full-wave simulations were performed using a commercial three-dimensional finite-difference time-domain (3D-FDTD) solver. A single unit cell was simulated using periodic boundary conditions in the in-plane dimensions and perfectly matched layers in the out-of-plane dimension. The structures were modelled using fully dispersive optical material properties for silica~\cite{Palik1985} and for gold~\cite{Johnson1972}. Minimal artificial absorption ($\mathrm{Im}(n)\,{\sim}\,10^{-4}$) was added to the background medium to reduce numerical divergences.

\paragraph*{LSA:} The lattice sum approach (LSA) is a variant of the discrete-dipole approximation (DDA) method~\cite{Draine1994}. It is a semi-analytic calculation method that has been found to produce accurate results for plasmonic arrays~\cite{Orad2019,Huttunen2016,Markel2005}. The main assumption in LSA when compared to DDA is that the dipole moments of all interacting nanoparticles are assumed to be identical~\cite{Huttunen2016}. The main benefit of using LSA for our application compared to alternatives such as FDTD is its capability to model finite-sized arrays with an arbitrary number of nanostructures, by assuming that the overall response of the array closely follows the responses of the nanoparticles at the center of the array. By comparing simulations performed using the LSA against the DDA, this assumption has also been found to be quite accurate~\cite{Huttunen2016}. Its rapid simulation time makes it a useful tool for iterating many simulations to study trends and behaviours of entire metasurfaces, especially for finite array effects, such as the effect of array size on the $Q$-factor.

Using the LSA approach, the dipole moment $\vec{p}$ of any particle in the array is written as  
\begin{equation} \label{Eq:p_lsa}
\vec{p} = \frac{\epsilon_0 \alpha(\omega) \vec{E}_\mathrm{inc}}{1 - \epsilon_0\alpha(\omega) \mathscr{S}(\omega)} \equiv \epsilon_0 \alpha^*(\omega) \vec{E}_\mathrm{inc}, 
\end{equation}
where the effect of inter-particle coupling is incorporated in the lattice sum $\mathscr{S}$, and $\alpha^*$ is the effective polarizability. This equation produces Eq.~\eqref{EQ:alpha_star} in the main text. The calculations presented in this work also incorporate a modified long-wavelength correction~\cite{Jensen1999}:
\begin{equation}
    \alpha(\omega) \rightarrow \frac{ \alpha_\mathrm{static}(\omega)}{1-\frac{2}{3}ik^3\alpha_\mathrm{static}(\omega)-\frac{k^2}{l}\alpha_\mathrm{static}(\omega)},
\end{equation}
where $k$ is the wavenumber in the background medium $k=(2\pi n/\lambda)$ and $l$ is the effective particle radius. Also here, minimal artificial absorption ($\mathrm{Im}(n)=6 \times 10^{-4}$) was added to the refractive index $n=1.452$ of the background medium to reduce numerical divergences associated with the approach when considering large arrays~\cite{Markel2005}. We set $l=180$~nm for all calculations. The static polarizability of the nanoparticle is given by
\begin{equation}
    \alpha_\mathrm{static}(\omega) = \frac{A_0}{\omega-\omega_0-i\gamma},
\end{equation}
where $A_0$ is the oscillator strength, $\omega_0=2\pi c / \lambda_\mathrm{LSPR}$ corresponds to the nanoparticle resonance frequency and $\gamma$ is the damping term.

For a planar array of $N$ dipoles, the lattice sum term $\mathscr{S}$ is
\begin{align} \label{Eq:S_in} 
\mathscr{S}(\omega) = \sum\limits^N_{j=1} \frac{\mathrm{exp}(\mathrm{i}kr_j)}{\epsilon_0 r_{j}} \Bigg[ k^2 +\frac{(1-\mathrm{i}kr_j )(3 \cos^2\theta_{j}-1)}{r_j^2} \Bigg],  
\end{align}
where $r_j$ is the distance to the $j^{\mathrm{th}}$ dipole, and $\theta_j$ is the angle between $\vec{r}_j$ and the dipole moment $\vec{p}$.

The optical transmission spectra can be obtained by using the optical theorem, $ \mathrm{Ext} \propto k \mathrm{Im}(\alpha^*)$~\cite{Jackson}:
\begin{align}
T(\omega)=1-\frac{4\pi k}{P_x P_y}\mathrm{Im}[\alpha^*(\omega)],\label{EQ:T}
\end{align} 
where $P_x$ and $P_y$ are the lattice constants along the $x$ and $y$ dimensions, respectively.

To produce the plots in Fig.~\ref{fig:schematic}d, we performed an LSA calculation using the following parameters for the single dipole: $\lambda_\mathrm{LSPR} = 780$~nm; $A_0 = 3.46\times10^{-7}\mathrm{~m}^3/\mathrm{s}$, $\gamma = 8.5\times10^{13} \mathrm{~s}^{-1}$. LSA parameters were determined by matching to FDTD data. The lattice constants were $P_x = 500$~nm and $P_y=1067.5$~nm. The total array size was $600\times600$~\micron$^2$, corresponding to $N_x=1200 \times N_y =562$~nanostructures, respectively. The LSA calculations in Fig.~\ref{FIG:ArraySize} used these same parameters, but varied the total number of nanostructures.

To calculate the figures in Fig.~\ref{FIG:CriticalCoupling}a~--~c, we performed a series of LSA calculation using the following parameters for the particle: $A_0 = 3.98\times10^{-7}\mathrm{~m}^3/\mathrm{s}$, $\gamma = 1/[2\pi(2.1\mathrm{~fs})]\approx 7.6\times10^{13}\mathrm{~s}^{-1}$. The dipole resonance wavelengths $\lambda_\mathrm{LSPR}$ were 800, 833, 866, 900, 933, 966 and 1000~nm, respectively. Based on performed FDTD simulations, these resonance wavelengths could correspond to rectangular gold nanostructures with widths of $L_x=110$, 120, 130, 140, 150, 160, and 170~nm, respectively, if $L_y = 190$~nm, and $t = 20$~nm. (Note that in the main text, $L_y = 200$~nm). See \SI{SEC:particle_dimension_Sweep} for the corresponding simulations. The lattice constants were $P_x=500$~nm and $P_y=1060$~nm, respectively. The total array size was $600\times 600$~\micron$^2$, corresponding to $N_x=1200 \times N_y =567$~nanostructures, respectively. To obtain Fig.~\ref{FIG:CriticalCoupling}d, a series of LSA calculations were performed for many values of $\lambda_\mathrm{LSPR}$ ranging from 800~nm to 1000~nm, and the $Q$-factors were extracted from the results using a fit to a Lorentzian. The curves in \textbf{(d)} come from repeating this procedure with oscillator strengths of $A_0 = 3.98\times10^{-7}$, $4.38\times10^{-7}$ and $4.77\times10^{-7}\mathrm{~m}^3/\mathrm{s}$.

\subsection*{Device details}
We fabricated different metasurface devices with array sizes of 300$\times$300~\micron{}$^2$, 400$\times$400~\micron{}$^2$, 500$\times$500~\micron{}$^2$, and 600$\times$600~\micron{}$^2$, with a corresponding number of participating nanostructures of 600$\times$284, 800$\times$378, 1000$\times$472, and 1200$\times$567, respectively. The lattice constants of the rectangular arrays are $P_x=500$~nm $\times$ $P_y=1060$~nm. The dimensions of the rectangular gold nanostructures are  $L_x=130$~nm $\times$ $L_y=200$~nm, with a thickness of $t=20$~nm. The lattice is embedded within a homogeneous background $n\approx1.46$.

\subsection*{Fabrication}
The metasurfaces are fabricated using a standard metal lift-off process. We start with a fused silica substrate. We deposit a silica undercladding layer using sputtering. We then define the pattern using electron-beam lithography in a positive tone resist bi-layer with the help of a commercial conductive polymer. The mask was designed using shape-correction proximity error correction~\cite{Schulz2015a} to correct for corner rounding. Following development, a thin adhesion layer of chromium (0.2 nm thick) is deposited using e-beam evaporation, followed by a layer of gold deposited using thermal evaporation. Lift-off is performed, and a final protective silica cladding layer is deposited using sputtering. The initial and final silica layers are sputtered using the same tool under the same conditions to ensure that the environment surrounding the metasurface is completely homogeneous. Before characterization, the surface of the device is then covered in index-matching oil. The backside of the silica substrate is coated with an anti-reflective coating to minimize substrate-related etalon fringes.

\subsection*{Characterization}
See \SI{SEC:exmpt_setup} for a schematic of the experimental setup.

\paragraph*{Coherent light measurements:}  To measure the transmission spectra, we flood-illuminated all of the arrays in the sample using a collimated light beam from a broadband supercontinuum laser source. The wavelength spectrum of the source ranges from $\lambda=470$ to $2400$~nm. The beam comes from normal incidence along the $z$-direction with light polarized in the $x$-direction. The incident polarization is controlled using a broadband linear polarizing filter. Light transmitted by the metasurface is then imaged by a $f=35$~mm lens, and a 100~\micron{} pinhole is placed in the image plane to select the desired array. The transmitted light is collected in a large core ($400$~\micron) multimode fiber and analyzed using an optical spectrum analyzer, and is normalized to a background trace of the substrate without gold nanostructures. The resolution of the spectrometer is set to 0.01~nm. 

\paragraph*{Incoherent light measurements:} Here, the experiment goes as above, but the samples are excited using a collimated tungsten-halogen light source (ranging from $\lambda=300$ to $2600$~nm) and a 400~\micron{} pinhole.

\vspace{1em}\noindent\textbf{Acknowledgements} Fabrication in this work was performed in part at the Centre for Research in Photonics at the University of Ottawa (CRPuO). The authors acknowledge support from the Canada Excellence Research Chairs (CERC) Program, the Canada Research Chairs (CRC) Program, and the Natural Sciences and Engineering Research Council of Canada (NSERC) Discovery funding program. MSBA acknowledges the support of the Ontario Graduate Scholarship (OGS), the University of Ottawa Excellence Scholarship, and the University of Ottawa International Experience Scholarship. OR acknowledges the support of the Banting Postdoctoral Fellowship of the NSERC. YM was supported by the Mitacs Globalink Research Award. MJH acknowledges the support of the Academy of Finland (Grant No. 308596) and the Flagship of Photonics Research and Innovation (PREIN) funded by the Academy of Finland (Grant No. 320165). 

\vspace{1em}\noindent\textbf{Author Contributions} MSBA, OR, and MJH conceived the basic idea for this work. OR and MSBA performed the FDTD simulations. MJH, MSBA and OR performed the lattice sum calculations. OR and GC fabricated the device. MZA and MJH designed the preliminary experimental setup. MSBA and YM carried out the measurements. OR, MSBA, and YM analysed the experimental results. JU, BS, JMM, MJH, RWB, and KD supervised the research and the development of the manuscript. MSBA and OR wrote the first draft of the manuscript. All co-authors subsequently took part in the revision process and approved the final copy of the manuscript. Portions of this work were presented at the 2020 SPIE Photonics West conference in San Francisco, CA.

\clearpage

\onecolumn
\renewcommand\thepage{S\arabic{page}} 
\setcounter{page}{1}
\renewcommand\thesection{S\arabic{section}} 
\setcounter{section}{0}
\renewcommand\thefigure{S\arabic{figure}}   
\setcounter{figure}{0}  
\renewcommand\theequation{S\arabic{equation}} 
\setcounter{equation}{0}

{\Huge Supplementary Information}
\vspace{1em}

\noindent Below is the supplementary information for \emph{Ultra-high-$Q$ resonances in plasmonic metasurfaces} by M. Saad Bin-Alam, Orad Reshef, Yaryna Mamchur, M. Zahirul Alam, Graham Carlow,  Jeremy Upham, Brian T. Sullivan, Jean-Michel M\'{e}nard, Mikko J. Huttunen, Robert W. Boyd, and Ksenia Dolgaleva.
In Sec.~\ref{SEC:fit}, we present supporting material for Fig.~\ref{FIG:CriticalCoupling}d and Fig.~\ref{FIG:ArraySize}b. In Sec.~\ref{SEC:SLR_Type} we determine the type of SLR by looking directly at the polarizability and the lattice sum.  Section~\ref{SEC:particle_dimension_Sweep} shows the dependence of the LSPR and SLR behaviours on the particle geometry, produced using FDTD simulations. It also contains additional measurement results for a different metasurface with the same lattice geometry. In Sec.~\ref{SEC:device_image}, we present a representative image of a fabricated device. In Sec.~\ref{SEC:exmpt_setup}, we describe our experimental setup. 

\section{$Q$-factor extraction}\label{SEC:fit}
\begin{figure}[H]
     \centering \includegraphics[width=1\linewidth]{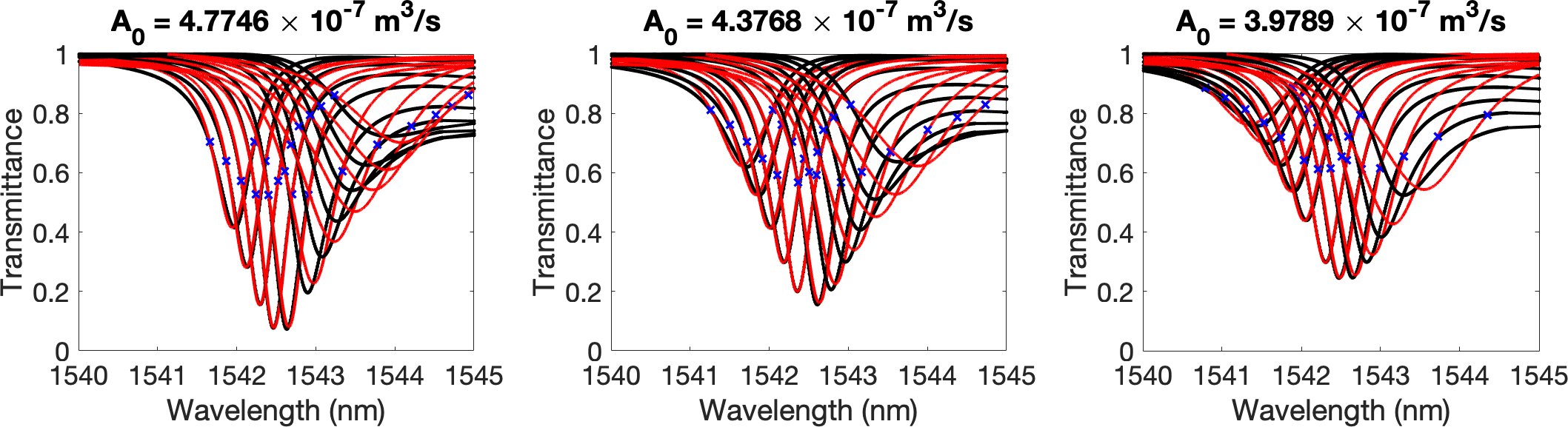}
     \caption{{\bf~|~Parameter sweeps of LSA calculations.} The black curves are calculated using the LSA using the values described in the Methods. The red curves correspond to Lorentzian fits.}
     \label{Fig:fits}
\end{figure}

\begin{figure}[H]
    \centering
    \subfloat[]{{ \includegraphics[width=0.45\linewidth]{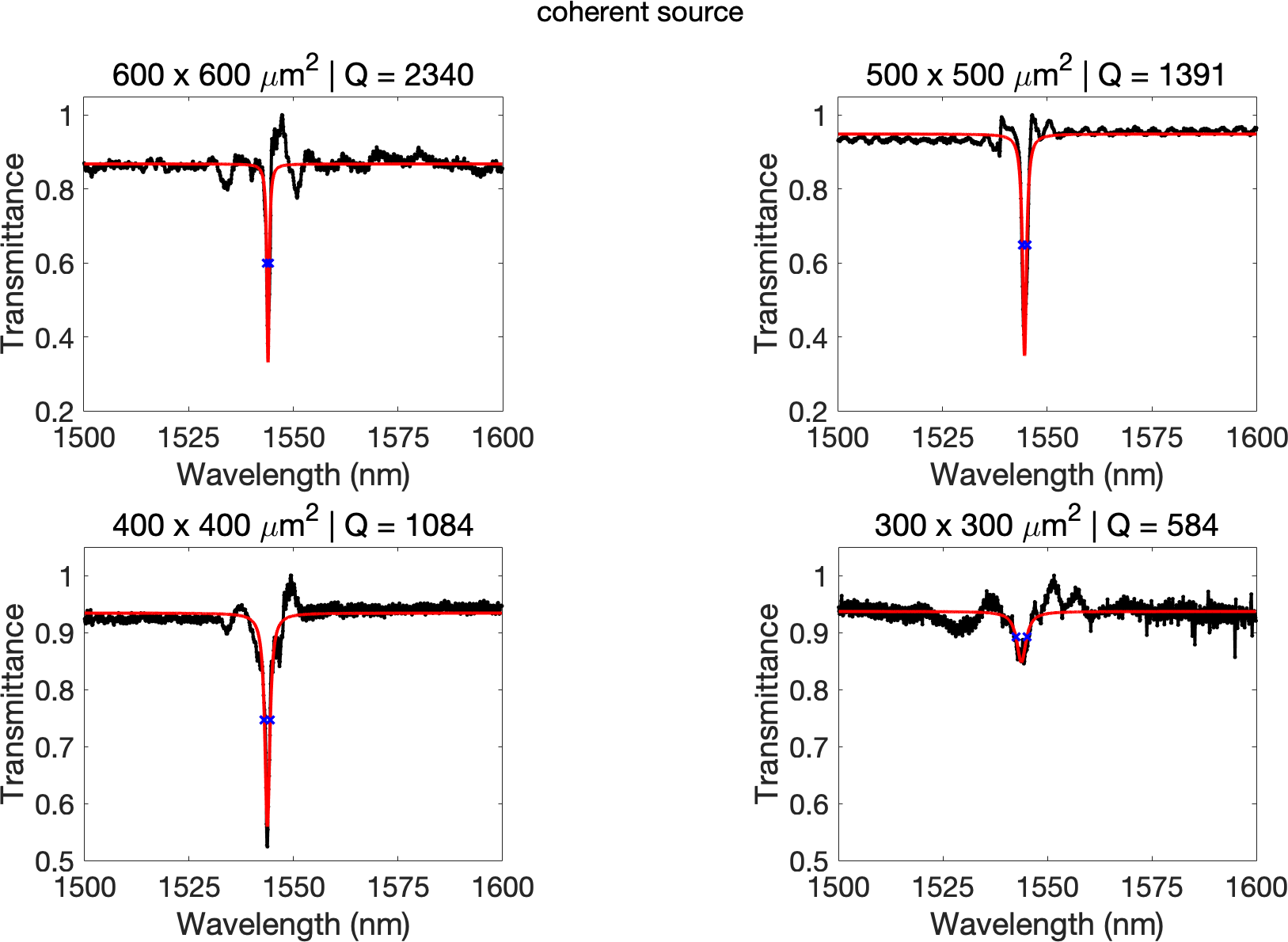} }}
    \qquad
    \subfloat[]{{ \includegraphics[width=0.45\linewidth]{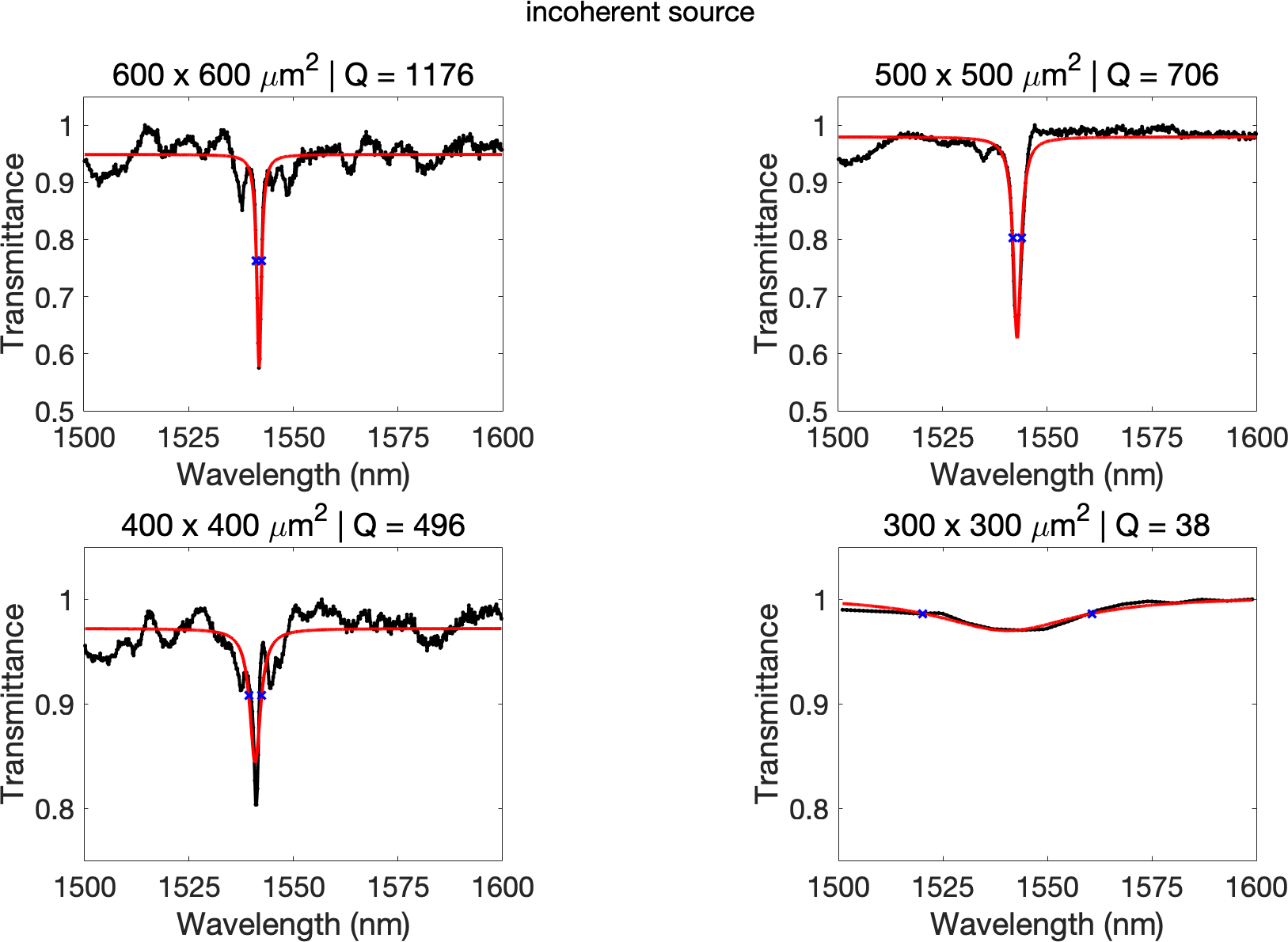} }}
    \caption{{\bf~|~Measurements of devices.} Measurements (black curves) of the devices described in the main text taken using {\bf a,} a coherent and {\bf b,} an incoherent source. The red lines correspond to Lorentzian fits. The array sizes and extracted $Q$-factors are indicated on the individual figures.}
    \label{FIG:fits_measurement}
\end{figure}

Figure~\ref{Fig:fits} shows Lorentzian fits to a series of LSA calculations with varying $\lambda_\mathrm{LSPR}$ (see Methods for values). The $Q$-factors extracted from these fits are used to produce the black curves in Fig.~\ref{FIG:CriticalCoupling}d. In Fig.~\ref{FIG:fits_measurement}, we reproduce the fits to the measurements that produced the values for Fig.~\ref{FIG:ArraySize}b.

\section{SLR Type}\label{SEC:SLR_Type}
\begin{figure}[H]
\centering
\includegraphics[width=90mm]{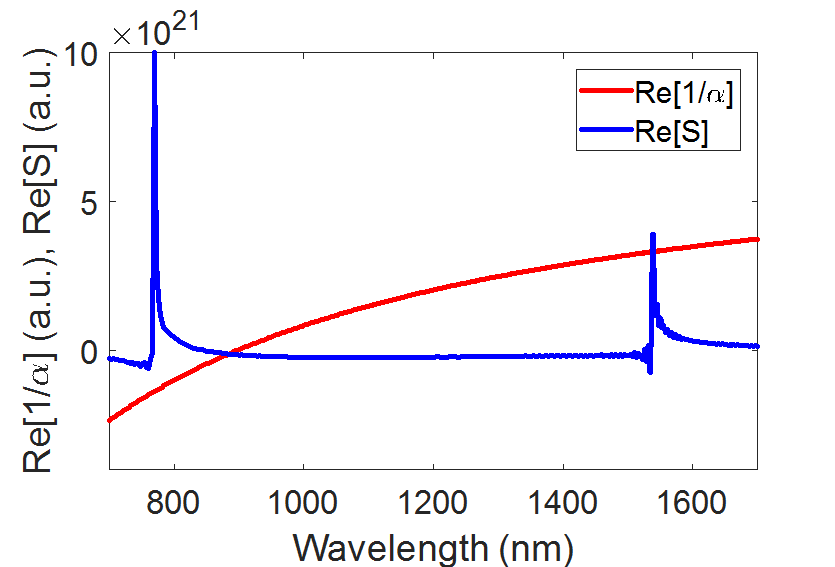}
\caption{{\bf~|~SLR type.} The real part of the inverse of the particle polarizability $\mathrm{Re}[1/\alpha]$ as well as the real part of the lattice sum $\mathrm{Re[\mathscr{S}]}$ for the metasurface in Fig.~\ref{fig:schematic}.}
\label{fig:Re[S]}
\end{figure}

Figure~\ref{fig:Re[S]} shows the real part of the inverse of the particle polarizability $\mathrm{Re}[1/\alpha]$ as well as the real part of the lattice sum $\mathrm{Re[\mathscr{S}]}$ for the metasurface in Fig.~\ref{fig:schematic}. As these two values cross twice near $\lambda_\mathrm{SLR}$, this SLR is considered to be an SLR of the first type according to the nomenclature of Ref.~\cite{Kravets2018}.

\section{Dependence of SLR behaviour on particle dimensions}\label{SEC:particle_dimension_Sweep}

\begin{figure}[H]
\centering
\includegraphics[width=1\linewidth]{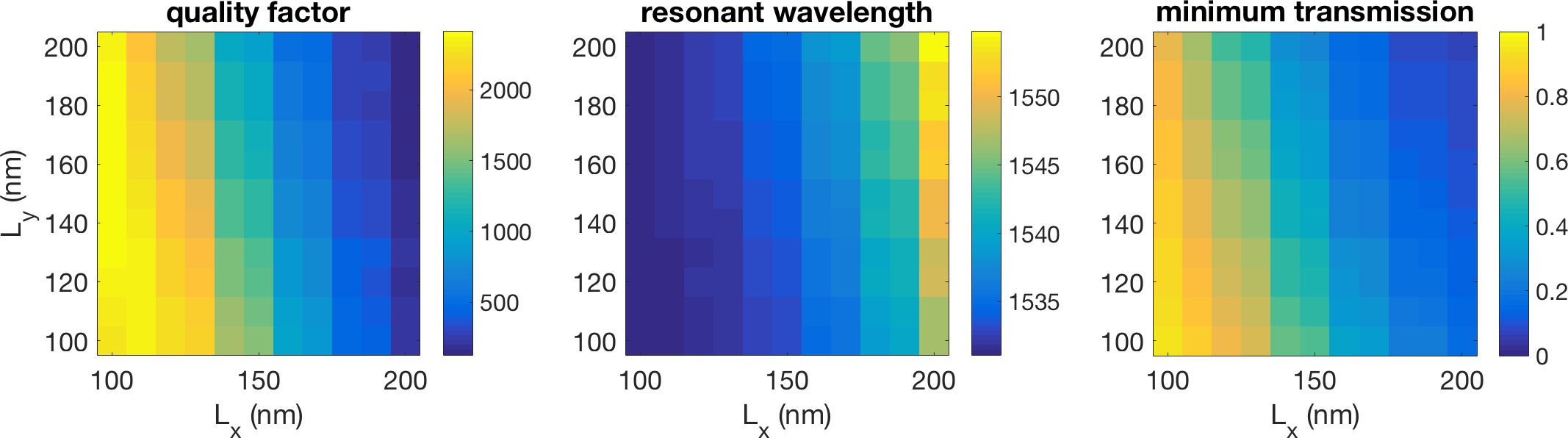}
\caption{{\bf~|~Particle dimension sweep.} Quality factor $Q$ (left), resonant wavelength $\lambda_\mathrm{SLR}$ (center), and minimum transmission as a function of particle dimensions $L_x$ and $L_y$, extracted from full-wave simulations performed with FDTD.}
\label{fig:dimension_sweep}
\end{figure}

To explicitly demonstrate how changing the dimensions of the nanoparticle may affect the properties of the SLR, we perform full-wave simulations in FDTD using a series of particle geometries. Figure~\ref{fig:dimension_sweep} depicts the simulation results. Not only the $Q$-factor, but also $\lambda_\mathrm{SLR}$ and the extinction ratio are all affected by changes in the particle dimensions.

\begin{figure}[H]
\centering
\includegraphics[width=0.40\linewidth]{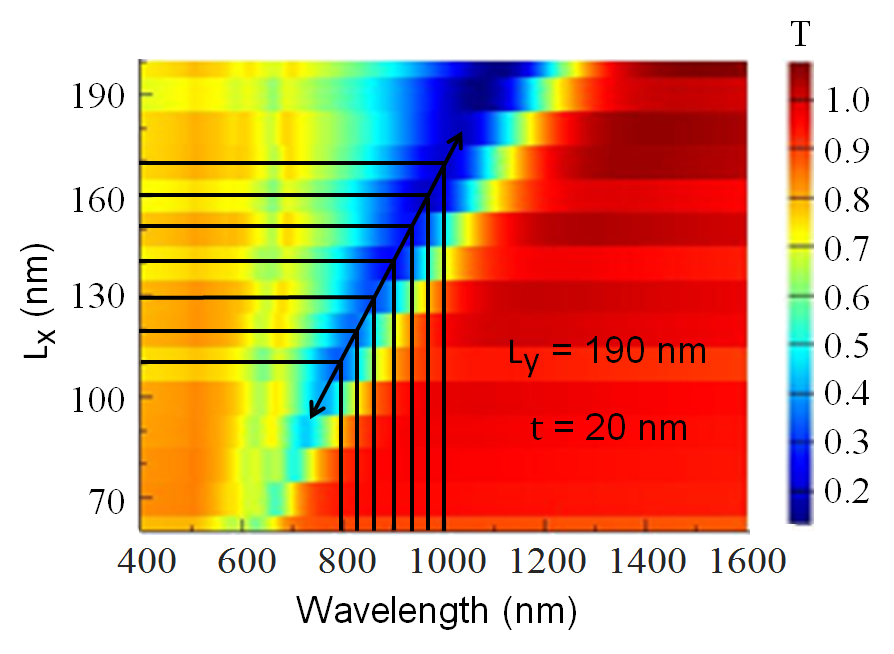}
\caption{{\bf~|~Particle dimension vs. LSPR wavelength.} $\lambda_\mathrm{LSPR}$ (shown in $x$-axis) linearly increases alongside the particle length $L_x$ towards the light polarization (shown in $y$-axis). This relation is extracted from full-wave simulations performed with FDTD.}
\label{fig:dimension_LSPR}
\end{figure}

\begin{figure}[H]
\centering
\includegraphics[width=92.5mm]{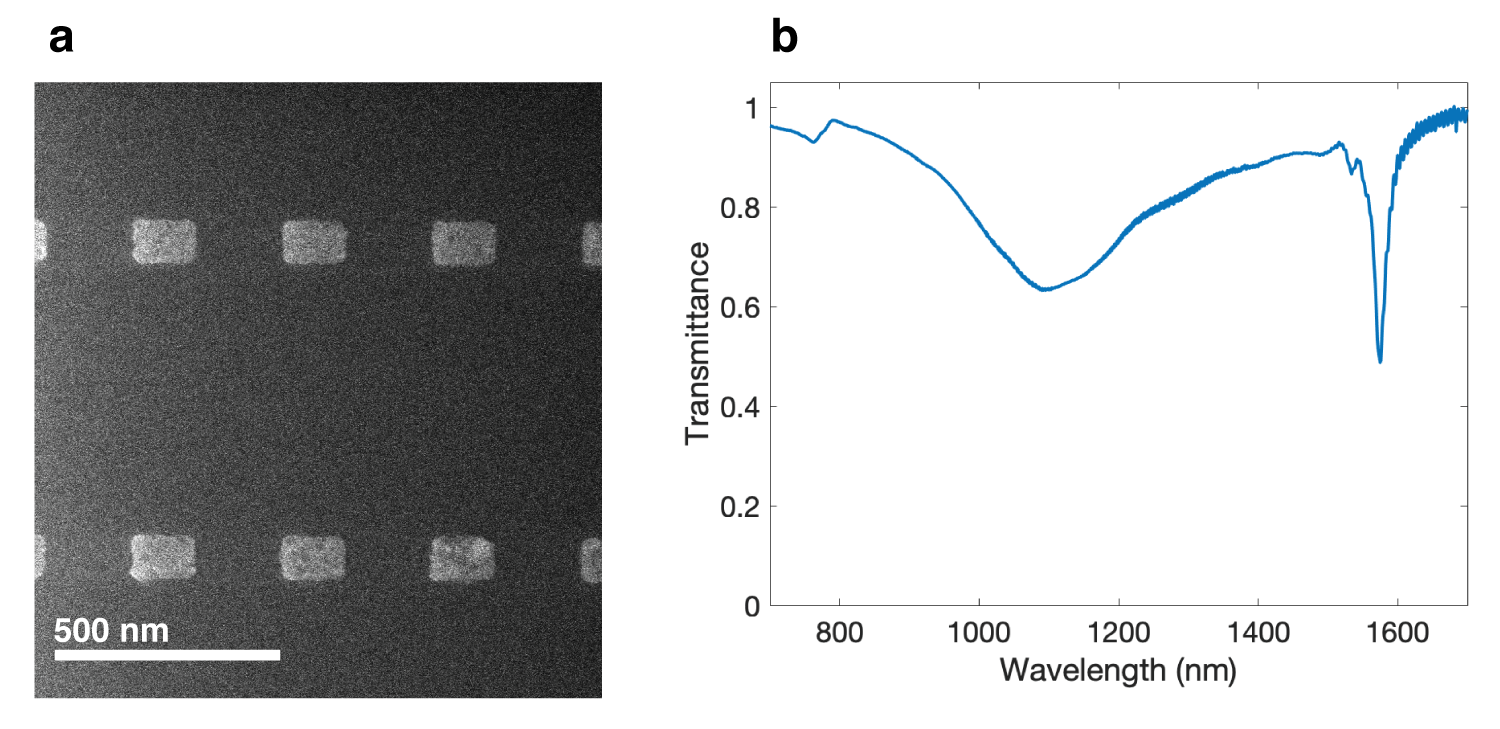}
\caption{{\bf~|~Metasurface with same lattice constants but rotated nanoparticles.} This device consists of a $400\times 400$~\micron$^2$ array with $L_x=200$~nm, $L_y=130$~nm, $P_x=500$~nm and $P_y=1060$~nm. The SLR is also located at $\lambda_\mathrm{SLR}=1550$~nm, but here $Q=80$.}
\label{FIG:OtherOrientation}
\end{figure}

In Fig.~\ref{FIG:OtherOrientation}a, we present a different $400 \times 400$~\micron$^2$ array with a nanoparticle geometry of $L_x = 200$~nm, $L_y=130$~nm. Note that the nanoparticle dimensions are identical to those presented in the arrays above, but due to being rotated by 90 degrees, their resulting particle polarizabilities are completely different. The lattice constants are identical to the arrays presented in the main text, that is, $P_x = 500$~nm and $P_y=1060$~nm. 

The measurements in Fig.~\ref{FIG:OtherOrientation}b were performed using an incoherent source. Here, due to the different polarizability, $\lambda_\mathrm{LSPR}$ is red-shifted (1100 nm vs 840 nm), and consequently, the SLR is dramatically affected: in comparison to the matching array in the main text which has an SLR of $Q=500$, the SLR here only has a $Q=80$, despite having the same lattice constants and nanoparticle geometries. This further demonstrates the importance of the polarizability to the $Q$ of the SLR.

\section{Image of the device}\label{SEC:device_image}
\begin{figure}[H]
\centering
\includegraphics[width=50mm]{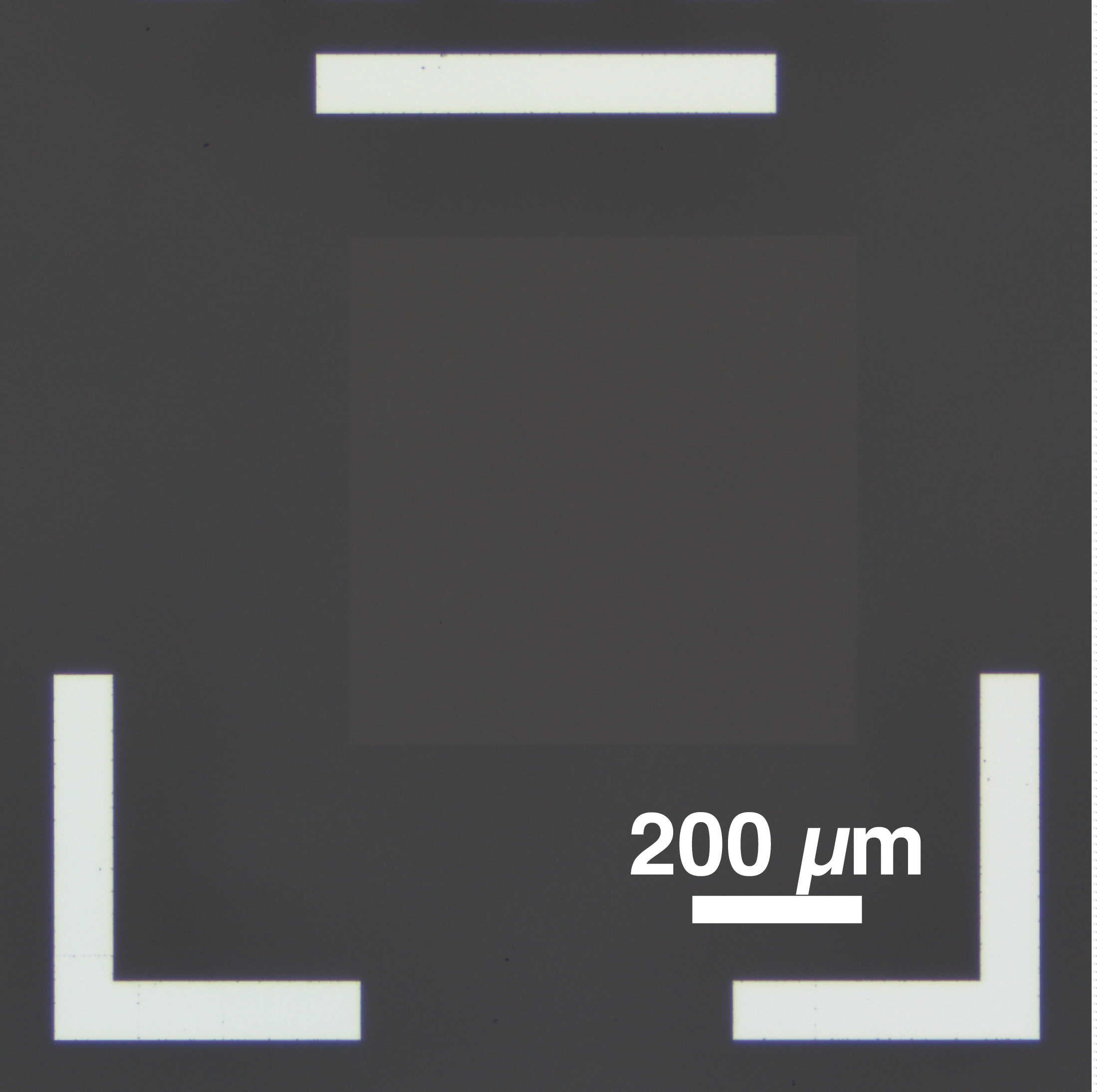}
\caption{{\bf~|~Large-area high-$\mathbf{Q}$ SLR device.} Optical image of a $600\times 600$~\micron$^2$ array.}
\label{fig:opticalimage}
\end{figure}

Figure~\ref{fig:opticalimage} shows a typical optical image for one of the devices taken with a bright field microscope. Surrounding the device are large aluminum alignment marks to help locate the device in the experimental setup.

\section{Experimental setup}\label{SEC:exmpt_setup}
A broadband source is collimated and is polarized using a broadband linear polarizing filter. A first iris is optionally placed to help align the sample in the center of the beam. The beam is then passed through the sample. The surface of the device is imaged using an $f_2=35$~mm lens, and a pinhole is placed in the image plane to select the desired array. The transmitted light is collected in a large core (400~\micron~diameter) multimode fiber and is analyzed using an optical spectrum analyzer.

\begin{figure}[H]
\centering
\includegraphics[width=140mm]{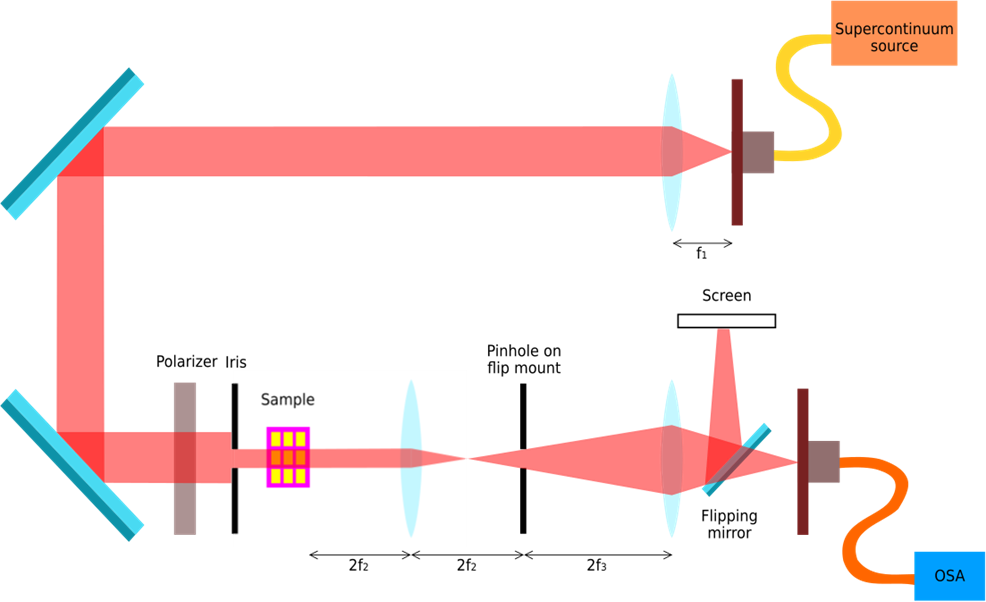}
\caption{{\bf~|~Experimental setup.} The metasurface is excited by a broadband collimated and polarized beam. Light is collected from the image plane of the metasurface and detected using a camera or a spectrum analyser.}
\label{FIG:expmt}
\end{figure}

\end{document}